\documentclass[twocolumn,aps,pra,superscriptaddress,amsfonts]{revtex4-1}

\usepackage{amsmath}
\usepackage{graphicx}
\usepackage{bm}
\usepackage{array}
\usepackage{dcolumn}
\usepackage{hepparticles}
\usepackage{heppennames}
\usepackage{hepnicenames}
\usepackage{epstopdf}

\begin{document}

\newcommand{\ket}[1]{{| {#1} \rangle}}
\newcommand{\bra}[1]{{\langle {#1}|}}

\newcommand{\FigureOne}{
\begin{figure}[htbp!]
\includegraphics*[width=\columnwidth]{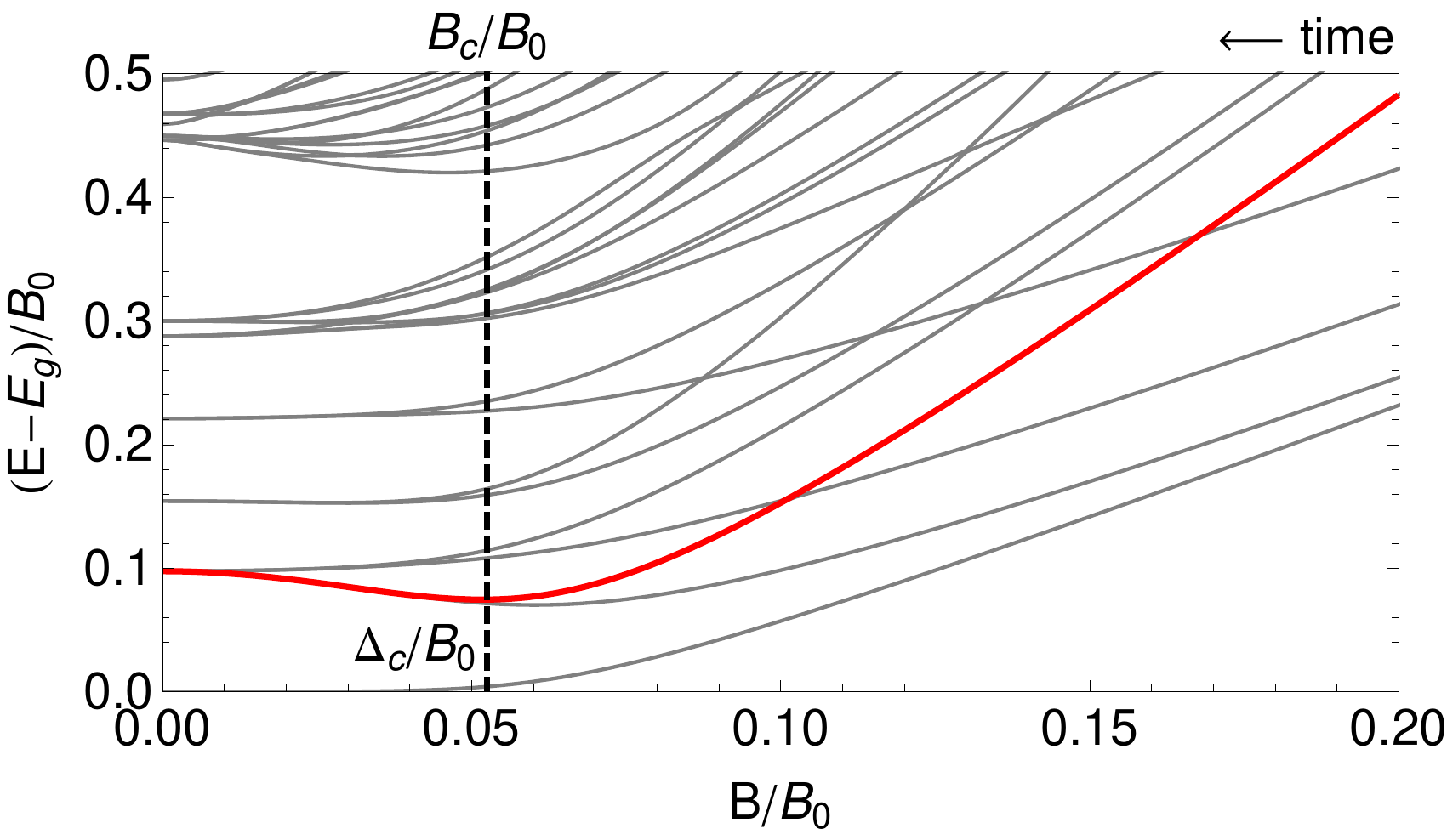}
\caption{(Color Online) Low-lying energy eigenvalues of Eqn. \ref{eqn:Hamiltonian} for $N=6$, with the ground state energy $E_g$ set to 0, $B_0=5J_{\text{max}}$, and the long-range $J_{i,j}$ couplings determined from experimental conditions (see text). Indicated in bold red is the first coupled excited state, the minimum of which determines the critical field $B_c$ and the critical gap $\Delta_c$.}
\label{fig:EnergyLevels}
\end{figure}
}

\newcommand{\FigureTwo}{
\begin{figure}[t!]
\includegraphics*[width=\columnwidth]{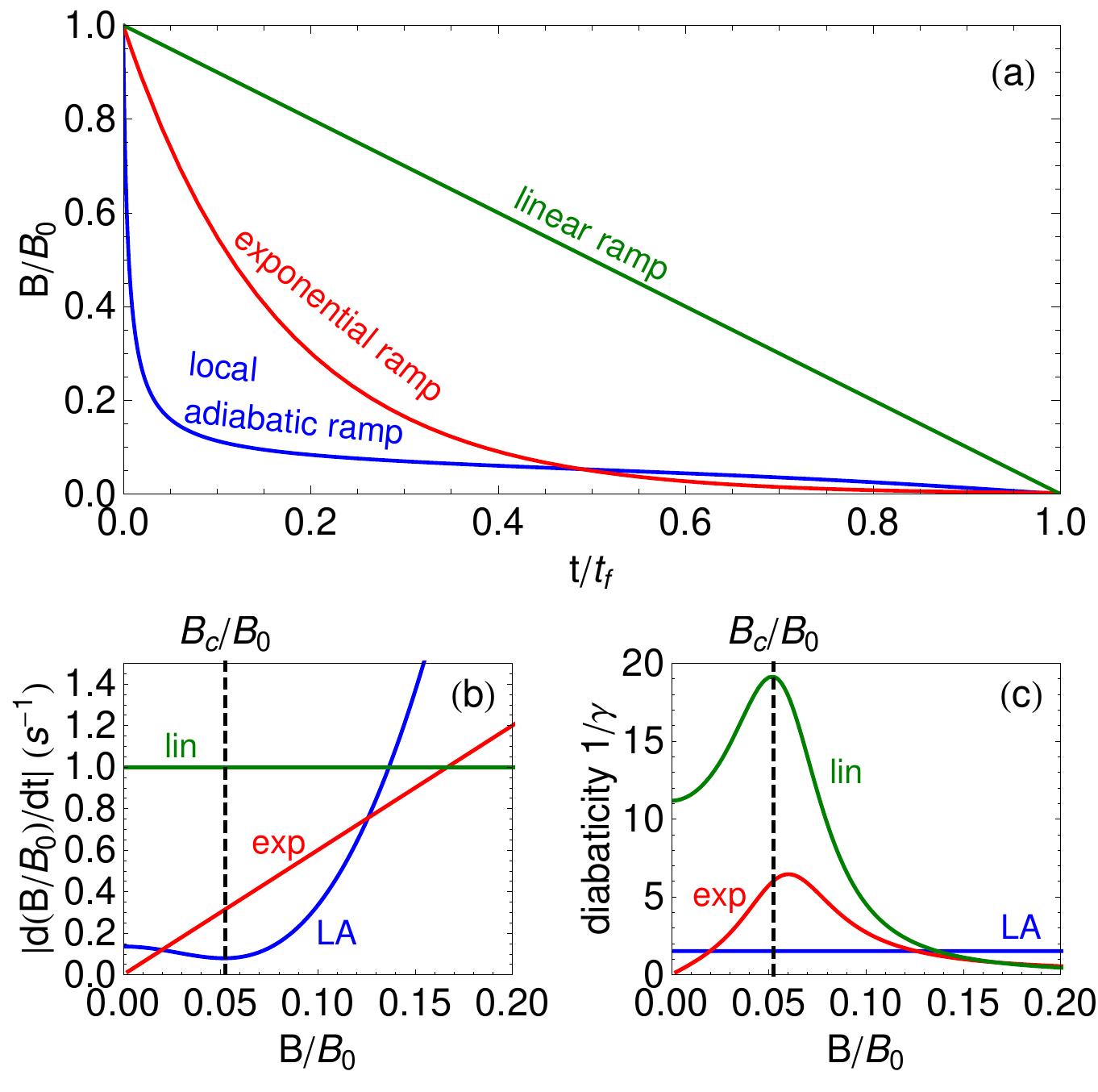}
\caption{(Color Online) (a) Local adiabatic ramp profile calculated for the energy levels in Fig. \ref{fig:EnergyLevels}, along with a linear ramp and an exponential ramp with decay constant $\tau=t_f/6$. (b) The slope of the local adiabatic (LA) ramp is minimized at the critical field value $B_c$, and is smaller than the slopes of the exponential and linear ramps at the critical point. (c) The inverse of the adiabaticity parameter $\gamma$ (see text) is peaked near the critical point for exponential and linear ramps but constant for the local adiabatic profile.}
\label{fig:rampprofiles}
\end{figure}
}

\newcommand{\FigureThree}{
\begin{figure}[h]
\includegraphics*[width=\columnwidth]{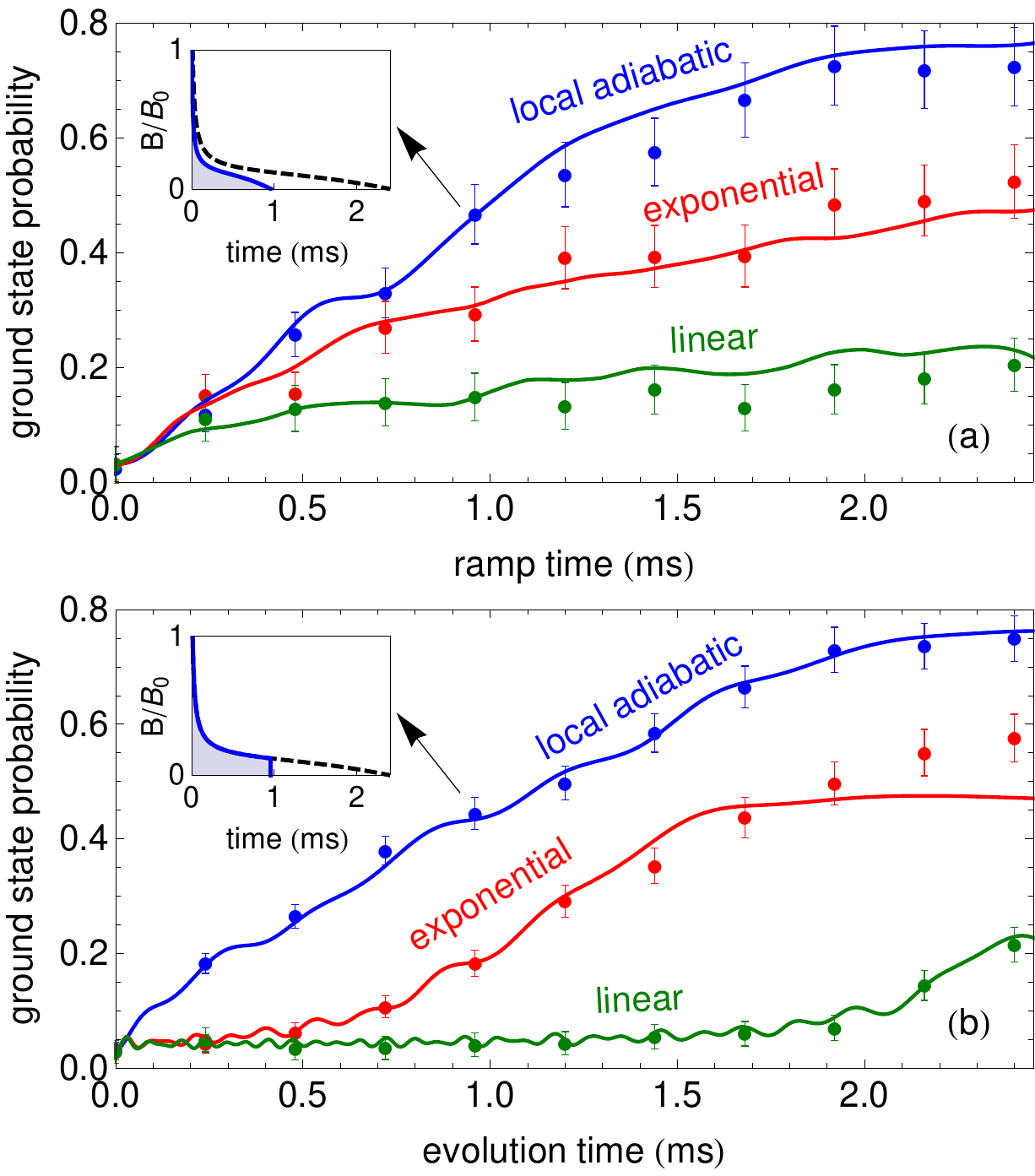}
\caption{(Color Online)  (a) Probability of preparing the AFM ground state after local adiabatic, exponential, and linear ramps with $t_f$ varied from 0 to 2.4 ms. The local adiabatic ramp gives the ground state with highest probability. Solid lines indicate the theoretical prediction. Inset: 0.96 ms local adiabatic ramp profile compared to the 2.4 ms profile (dotted). (b) Probability of preparing the AFM ground state for various times during $t_f=2.4$ ms simulations with three different ramp profiles. The linear ramp takes $\sim2.3$ ms to reach the critical point, while the local adiabatic and exponential ramps need only 1.2 ms. The inset shows the 2.4 ms local adiabatic profile evolved for 0.96 ms.}
\label{fig:rampdata}
\end{figure}
}

\newcommand{\FigureFour}{
\begin{figure}[t!]
\includegraphics*[width=\columnwidth]{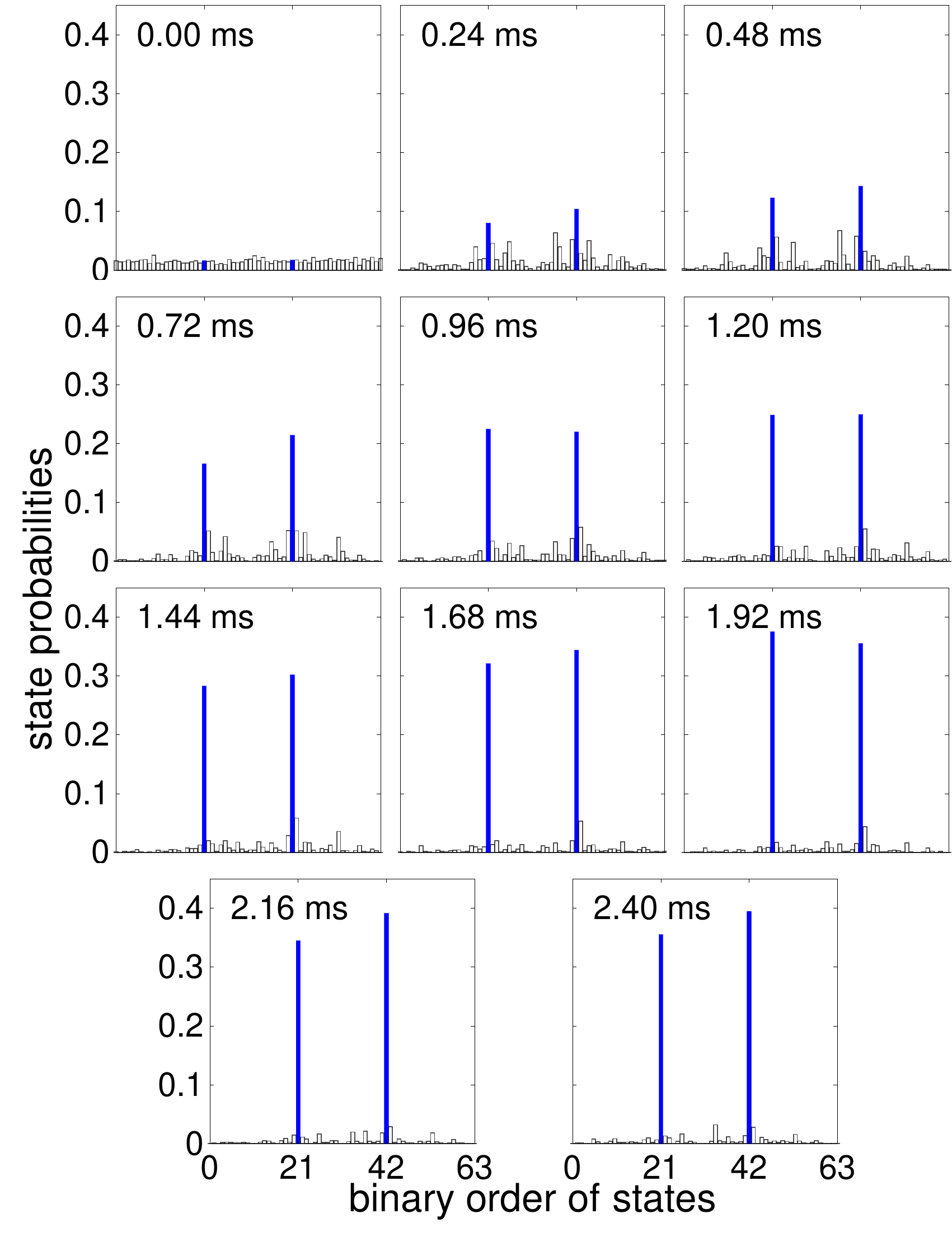}
\caption{(Color Online) State probabilities of all $2^6=64$ spin configurations for each local adiabatic data point in Fig. \ref{fig:rampdata}(a), ordered in binary (e.g. $\ket{010101} = 21$ and $\ket{101010}=42$). The two degenerate AFM states (solid blue) are the most prevalent for all times.}
\label{fig:MostProbableState}
\end{figure}
}

\newcommand{\FigureFive}{
\begin{figure}[t!]
\includegraphics*[width=\columnwidth]{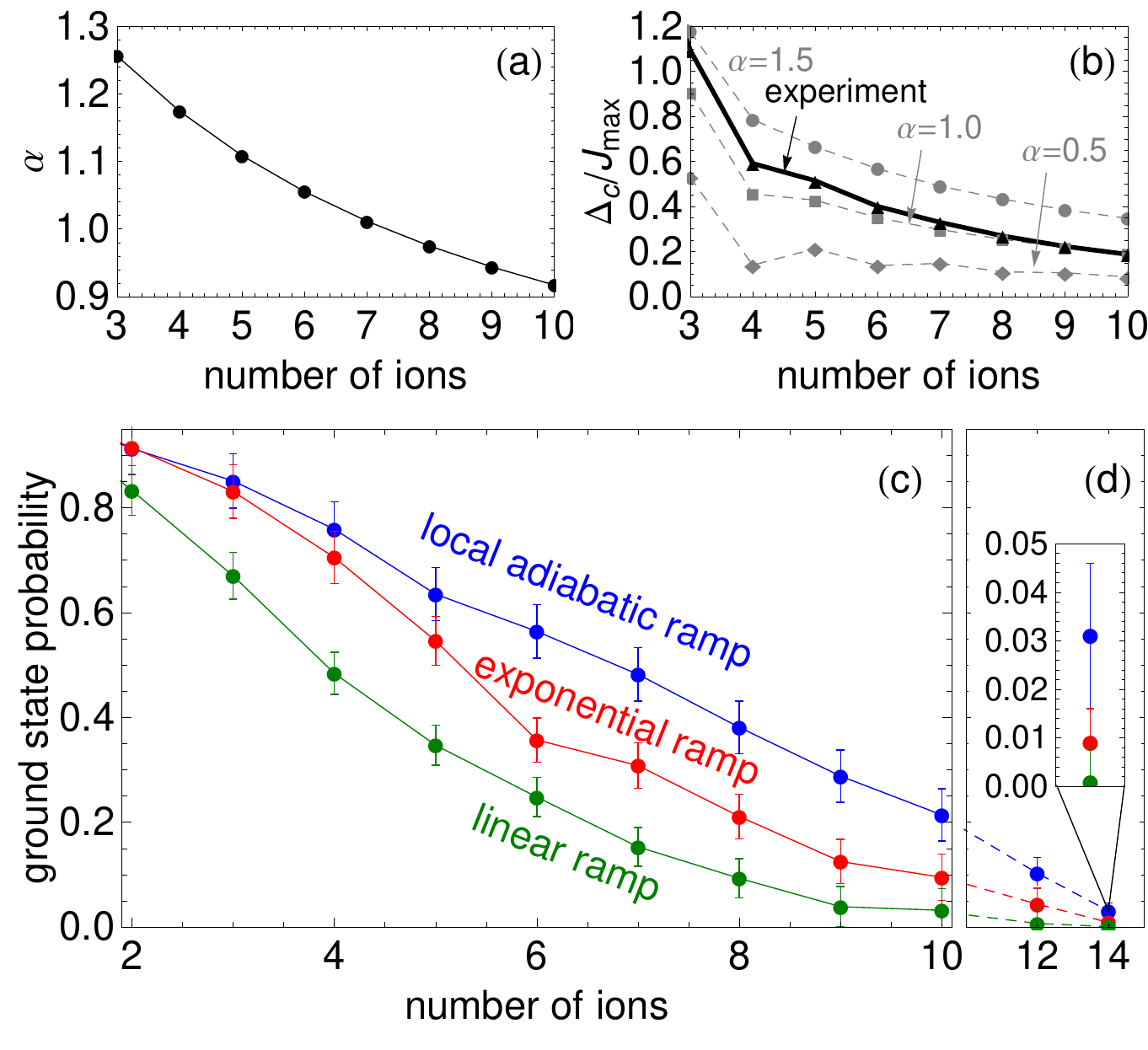}
\caption{(Color Online) (a) The long-range AFM interactions between spins $i$ and $j$ fall off as $\sim 1/|i-j|^\alpha$. For fixed trap voltages, increasing the number of ions leads to smaller $\alpha$ and longer-range interactions which increase frustration in the system. (b) The critical gap $\Delta_c$ between the ground and first coupled excited state shrinks for increasing $N$. The gap for experimental parameters is compared with three different curves that show the shrinking gap for fixed values of $\alpha$. (c) The measured ground state probability decreases with increasing $N$, reflecting the narrowing critical gap. Lines are to guide the eye. (d) An approximate local adiabatic ramp profile for 12 (14) ions yields a 10\% (3\%) probability of creating the ground state, much larger than the average state probability of 0.02\% (0.006\%).}
\label{fig:Niondata}
\end{figure}
}

\newcommand{\FigureSix}{
\begin{figure}[t!]
\includegraphics*[width=\columnwidth]{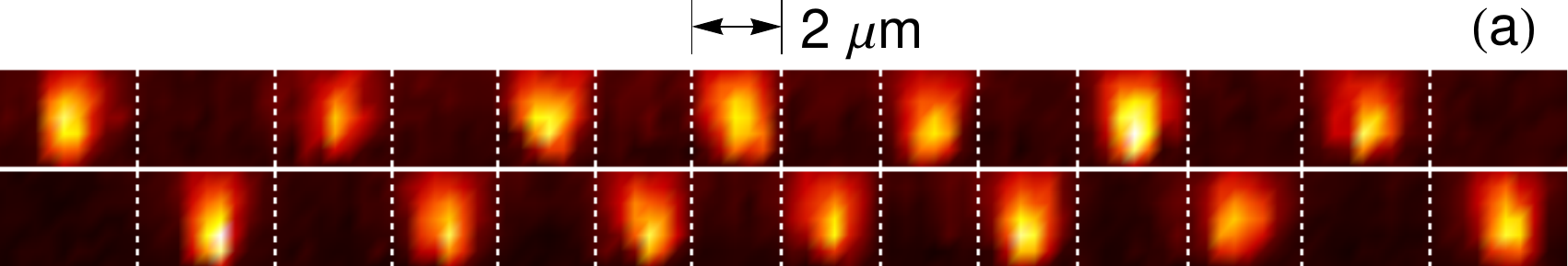}
\includegraphics*[width=\columnwidth]{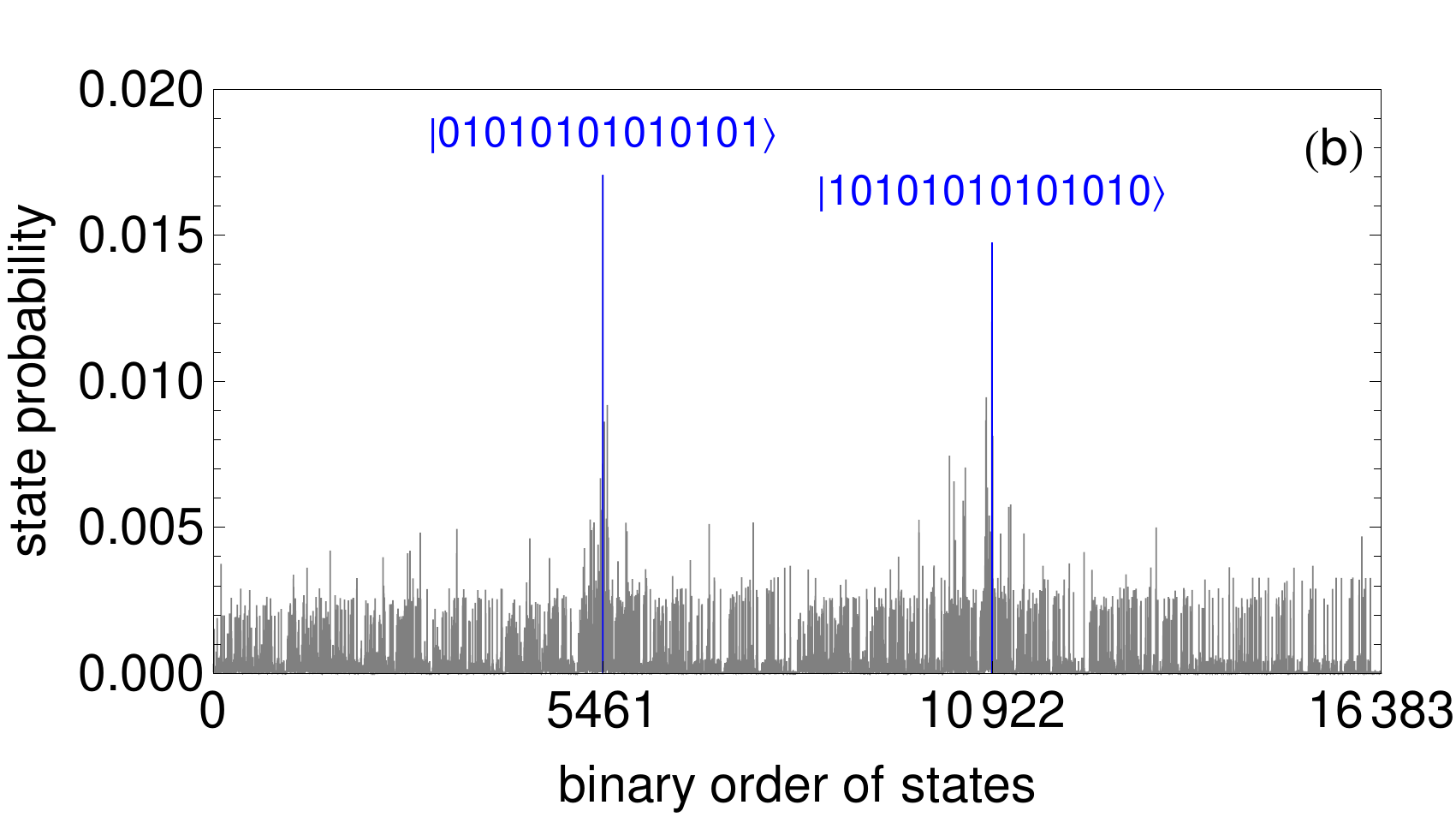}
\caption{(Color Online) (a) Camera images of experimentally prepared AFM ground states for $N=14$. (b) State probabilities of all $2^{14}=16384$ spin configurations for the 14-ion local adiabatic ramp in Fig. \ref{fig:Niondata}(d). The N\'eel-ordered ground states are unambiguously the most prevalent, despite a total probability of only 3\%.}
\label{fig:ProbsFourteen}
\end{figure}
}

\title{Experimental Performance of a Quantum Simulator: Optimizing Adiabatic Evolution and Identifying Many-Body Ground States}


\author{P. Richerme, C. Senko, J. Smith, A. Lee, S. Korenblit, and C. Monroe}
\affiliation{Joint Quantum Institute, University of Maryland Department of Physics and National Institute of Standards and Technology, College Park, MD 20742}

\date{\today}

\begin{abstract}     
We use local adiabatic evolution to experimentally create and determine the ground state spin ordering of a fully-connected Ising model with up to 14 spins. Local adiabatic evolution -- in which the system evolution rate is a function of the instantaneous energy gap -- is found to maximize the ground state probability compared with other adiabatic methods while only requiring knowledge of the lowest $\sim N$ of the $2^N$ Hamiltonian eigenvalues. We also demonstrate that the ground state ordering can be experimentally identified as the most probable of all possible spin configurations, even when the evolution is highly non-adiabatic.

\end{abstract}

\maketitle

\section{Introduction}
The investigation of quantum many-body systems and energy optimization problems often begins with the preparation or characterization of the ground state. A number of classical methods can quickly find the ground state for a wide range of many-body problems \cite{DMRGReview,QMCReview,SandvikSSE,DynamicalMeanFieldReview,DensityFunctionalTheory}, and specialized techniques \cite{SuzukiTrotterI, SuzukiTrotterII} can be used to find the ground state of large systems in certain instances \cite{LargeIsingI,LargeIsingII}. However, the fully-connected Ising model is known to be NP-complete \cite{IsingNPComplete,IsingNPComplete2}, and the exponential scaling of the state space with the system size limits solutions of many systems to only $N\lesssim 30$ spins \cite{Lanczos,SandvikScaling,MassivelyParallelQCSim}.

Such scaling issues motivated Feynman and others to propose quantum simulation, where a well-controlled quantum system is used to simulate a quantum system of interest \cite{FeynmanQSIM, LloydQSIM}. When paired with ideas underlying adiabatic quantum computation \cite{FarhiScience}, quantum simulation becomes a powerful way to find a many-body ground state by preparing the system in the ground state of a trivial Hamiltonian, adiabatically switching to the Hamiltonian of interest, and measuring the resulting ground state.

Even when the ground state of a particular many-body Hamiltonian is already known, preparing such a state with high probability can be useful for studying entanglement or dynamical processes -- both of which are generally difficult to solve classically \cite{CiracZollerQSIMCommentary}. In recent quantum simulation experiments, it has been necessary to start with a well-prepared ground state in order to probe frustrated antiferromagnetism \cite{QSIMNature2010,QSIM2013Science} or tunneling dynamics \cite{BlochSingleSpinAddressing}. Similarly, studies of defect production during non-equilibrium phase transitions \cite{ZurekQPT}, thermalization in closed quantum systems \cite{NonequilibriumDynamicsReview}, and excitation spectra of many-body Hamiltonians will likely require the spin ordering to be initialized into the ground state before proceeding.

In this paper, we show how local adiabatic evolution can be used for improved preparation and determination of many-body ground states in a trapped-ion quantum simulator. Compared with other adiabatic methods, local adiabatic evolution \cite{RCLocalAdiabaticEvolution} yields the highest probability of maintaining the ground state in a system that is made to evolve from an initial Hamiltonian to the Hamiltonian of interest. Compared with optimal control methods \cite{GRAPE,KrotovBook}, local adiabatic evolution may require knowledge of only the lowest $\sim N$ eigenstates of the Hamiltonian rather than all $2^N$. Using local adiabatic evolution in a system of up to 14 fully-connected spins, we demonstrate optimized ground state preparation as well as a method to find the ground state spin ordering even when the evolution is non-adiabatic.

The paper is organized as follows: Sec. \ref{sec:AQS} discusses the principles behind adiabatic quantum simulation as applied to our experimental system. In Sec. \ref{sec:Implementation}, we describe our physical implementation of an effective many-body spin system and the methods by which we perform adiabatic quantum simulations. Secs. \ref{sec:preparing} and \ref{sec:determining} demonstrate how local adiabatic evolution can be used to improve both ground state preparation and characterization, while Sec. \ref{sec:scaling} shows the robustness of the technique when scaled up to larger spin systems. In Sec. \ref{sec:conclusion} we offer some concluding remarks.

\section{Adiabatic Quantum Simulation}
\label{sec:AQS}
Adiabatic quantum simulation \cite{FeynmanQSIM, LloydQSIM,BulutaQSIMReview} applies the methods of adiabatic quantum computation \cite{FarhiAQC,FarhiScience,AQCvsQC} to solve interesting and difficult quantum problems. To date, adiabatic quantum simulations have been performed on a variety of different platforms \cite{NaturePhysicsInsightQSIM}, studying diverse problems such as quantum phase transitions \cite{GreinerAFM,GreinerAFMII}, quantum magnetism \cite{FriedenauerQSIM,QSIMPRB2010,QSIMNatureComm2011}, and quantum chemistry \cite{NMRHydrogenQSIM}. For the remainder of this paper, we will consider adiabatic quantum simulation within the context of the transverse-field Ising model.

The system Hamiltonian is given by
\begin{equation}
H=\sum_{i<j} J_{i,j} \sigma_x^{(i)} \sigma_x^{(j)} + B(t) \sum_{i} \sigma_y^{(i)}
\label{eqn:Hamiltonian}
\end{equation}
where $J_{i,j}$ are the Ising coupling strengths between spins $i$ and $j$, $B(t)$ is the magnitude of a time-dependent transverse magnetic field, $\sigma_\alpha^{(i)}$ is the Pauli spin operator for spin $i$ along the $\alpha$ direction, and Planck's constant $h=1$. We set $J_{i,j}>0$ for all $i\neq j$ in our experiments to generate long-range antiferromagnetic (AFM) spin-spin couplings.

An ideal adiabatic quantum simulation begins by initializing the spins to point along the transverse magnetic field $B_0 \hat{y}$, with $B_0 \gg \text{Max}(J_{i,j})$, which to good approximation is the instantaneous ground state of Eqn. \ref{eqn:Hamiltonian} at $t=0$. After initialization, the transverse field $B(t)$ is then ramped adiabatically from $B(t=0)=B_0$ to $B(t=t_f)=0$, ensuring that the system remains in its instantaneous ground state during its evolution. At the conclusion of the ramp, the ground state spin ordering of the Ising Hamiltonian (first term in Eqn. \ref{eqn:Hamiltonian}) may be either directly read out or used as a starting point for further experiments.

\FigureOne

Fig. \ref{fig:EnergyLevels} shows the energy level spectrum for the Hamiltonian in Eqn. \ref{eqn:Hamiltonian} for $N=6$ spins. Since the Hamiltonian obeys $Z_2$ symmetry (as well as parity symmetry in the experiments), the ground state is coupled to only a subset of the excited energy eigenstates. The first coupled excited state, shown in red in Fig. \ref{fig:EnergyLevels}, displays a general property seen in most adiabatic quantum simulations -- namely, the existence of a critical gap $\Delta_c$ that is central to parameterizing the adiabaticity of a given ramp. We will now explore three possible ramp profiles for transforming from the initial Hamiltonian to the problem Hamiltonian and discuss their implications for adiabaticity and ground state preparation.

\subsection{Linear Ramps}
For a linear ramp, the time-dependent transverse field $B(t)$ in Eqn. \ref{eqn:Hamiltonian} takes the form $B_{lin}(t)=B_0(1-t/t_f)$, with a ramp profile shown in Fig. \ref{fig:rampprofiles}(a). To determine whether such a ramp is adiabatic or not, we must compare to the adiabatic criterion \cite{MessiahQM}
\begin{equation}
\label{eqn:adiabatic}
\left|\frac{\dot{B}(t) \epsilon}{\Delta_c^2}\right| \ll 1
\end{equation}
where $\dot{B}(t)$ is the rate at which the transverse field is changed and 
$\epsilon=\text{Max}[\bra{e}dH/dB\ket{g}]$ is a number of order unity that parametrizes the coupling strength between the ground state $\ket{g}$ and the first coupled excited state $\ket{e}$. Eqn. \ref{eqn:adiabatic} highlights that fast ramps and small critical gaps can greatly decrease adiabaticity.

To satisfy the adiabatic criterion, a linear ramp must proceed slowly enough so that the total time \mbox{$t_f \gg B_0/\Delta_c^2$}. For the $N=6$ Ising Hamiltonian shown in Fig. \ref{fig:EnergyLevels}, $B_0=3.9$ kHz and $\Delta_c=0.29$ kHz, giving the adiabaticity requirement \mbox{$t_f\gg46$ ms}. As we will see in Sec. \ref{sec:preparing}, this time is exceptionally long compared with a maximum ramp time of $2.4$ ms in our apparatus (to avoid decoherence effects). We therefore seek alternative ways to decrease $B(t)$ more quickly while maintaining adiabaticity.

\subsection{Exponential Ramps}
Decreasing the transverse field exponentially according to $B_{exp}(t)=B_0\exp(-t/\tau)$, with $t_f=6\tau$, can yield a significantly more adiabatic evolution than linear ramps for the same $t_f$. Fig. \ref{fig:EnergyLevels} shows that the instantaneous gap $\Delta$ between the ground and first coupled excited state is large at the beginning of the ramp and small only when $B$ approaches $0$. Exponential ramps exploit this gap structure by quickly changing $B(t)$ at first, then gradually slowing the rate of change as $t\rightarrow t_f$.

At the critical point of the Hamiltonian shown in \mbox{Fig. \ref{fig:EnergyLevels}}, $|\dot{B}_{exp}(t)|=0.3B_0/t_f$. Adiabaticity (Eqn. \ref{eqn:adiabatic}) then requires $t_f \gg 14.5$ ms, a factor of 3 less time than the requirement found for linear evolution. Note that the adiabaticity gains of exponential ramps can be realized whenever the critical gap occurs towards the end of the ramp ($B_c/B_0 < \tau/t_f$), which is generally the case for the Ising Hamiltonian (Eqn \ref{eqn:Hamiltonian}).

\subsection{Local Adiabatic Ramps}
Local adiabatic ramps seek to keep the adiabaticity fixed at all points along the evolution by adjusting $\dot{B}(t)$ based on the instantaneous gap $\Delta (B(t))$ \cite{RCLocalAdiabaticEvolution, QuenchEcho}. If we define the adiabaticity parameter
\begin{equation}
\label{eqn:gamma}
\gamma=\left|\frac{\Delta^2 (B)}{\dot{B}(t)}\right|
\end{equation}
then a local adiabatic ramp would follow the profile $B(t)$ that solves the differential equation \ref{eqn:gamma} with $\gamma$ fixed. Adiabaticity then requires $\gamma\gg 1$.

To solve Eqn. \ref{eqn:gamma}, it is necessary to know $\Delta (B)$ everywhere along the evolution. This requires knowledge of the first \textit{coupled} excited state of the $N$-spin Hamiltonian (Eqn. \ref{eqn:Hamiltonian}), which is always the $3^{\text{rd}}$ excited state at small $B$ and the $(N+1)^{\text{st}}$ excited state at large $B$. Determining the local adiabatic evolution profile therefore relies on calculation of only the lowest $\sim N$ eigenvalues, which is much more computationally approachable than direct diagonalization of a $2^N\times 2^N$ matrix \cite{Lanczos}.

\FigureTwo

For a local adiabatic ramp, the critical time $t_c$ may be calculated by integrating Eqn. \ref{eqn:gamma}. Since $\dot{B}(t)$ is negative throughout the evolution, we find
\begin{equation}
\label{eqn:tc}
t_c=\gamma\int_{B_c}^{B_0}\frac{dB}{\Delta^2 (B)}
\end{equation}
Similarly, we may calculate the total evolution time
\begin{equation}
\label{eqn:tf}
t_f=\gamma\int_0^{B_0}\frac{dB}{\Delta^2 (B)}
\end{equation}
which shows a linear relationship between the total time $t_f$ and the adiabaticity parameter $\gamma$. Satisfying the adiabaticity condition $\gamma \gg 1$ for the Hamiltonian in \mbox{Fig. \ref{fig:EnergyLevels}} implies $t_f \gg 3.6$ ms, a factor of 4 and 12 less time than exponential and linear ramps, respectively. The fact that local adiabatic evolution can lead to faster ramps while satisfying adiabaticity has been well-explored in Ref. \cite{RCLocalAdiabaticEvolution}, where it was shown that local adiabatic ramps could recover the quadratic speedup of Grover's quantum search algorithm. In contrast, it was found that linear ramps offer no improvement over classical search \cite{FarhiAQC}.

Fig. \ref{fig:rampprofiles}(a) compares a linear, exponential, and local adiabatic ramp profile for the Hamiltonian shown in \mbox{Fig. \ref{fig:EnergyLevels}}. The local adiabatic ramp spends much of its time evolution in the vicinity of the critical point, since the transverse field changes slowly on account of the small instantaneous gap. This is further illustrated in Fig. \ref{fig:rampprofiles}(b), which shows that at the critical point, the slope of the local adiabatic ramp is minimized and smaller than slopes of the exponential or linear ramps. As a result, the inverse adiabaticity $1/\gamma$ is peaked near the critical point for exponential and linear ramps, greatly increasing the probability of non-adiabatic transitions away from the ground state (see Fig. \ref{fig:rampprofiles}(c)). By design, the local adiabatic ramp maintains constant adiabaticity for all values of $B$ and does not suffer from large non-adiabaticities near $B_c$.

\section{Physical Implementation}
\label{sec:Implementation}
Adiabatic quantum simulations are realized by applying the Hamiltonian (Eqn. \ref{eqn:Hamiltonian}) to an effective spin-1/2 system encoded in a linear chain of trapped $^{171}$Yb$^+$ ions \cite{PorrasCiracQSIM}. For this work, between 2 to 14 ions are held in an rf Paul trap with an axial center-of-mass frequency $f_z=0.7$ MHz and transverse frequencies $f_x=4.8$ MHz and $f_y=4.6$ MHz. The Ising spin states $\ket{0}_z$ and $\ket{1}_z$ are represented by the ion hyperfine clock states $^2S_{1/2}\ket{F=0,m_F=0}$ and $\ket{F=1,m_F=0}$, respectively. These states are split by $\omega_S/2\pi=12.642819$ GHz in a background magnetic field of $\sim 5$ G that defines the quantization axis, and their near-insensitivity to Zeeman shifts allows us to measure spin coherence times of longer than 1 second with no magnetic shielding \cite{YbDetection}.

Experiments begin by cooling the ion motion to deep within the Lamb-Dicke regime and optically pumping to the state $\ket{000\ldots}_z$. The effective spins are then coherently rotated to point along the $y-$direction of the Bloch sphere, which is the approximate instantaneous ground state of the Hamiltonian (Eqn. \ref{eqn:Hamiltonian}) at $t=0$. After initialization, we turn on the Hamiltonian and ramp $B(t)$ down with the desired profile. At $t=t_f$ the $\hat{x}$ component of each spin is coherently rotated back onto the $\hat{z}$ axis of the Bloch sphere. Measurement proceeds by illuminating the ions with 369.5 nm laser light resonant with the cycling $^2S_{1/2}$ to $^2P_{1/2}$ transition and imaging the spin-dependent fluorescence onto an intensified CCD camera \cite{QSIM2013Science}.

To apply the spin-spin interactions (first term of the Hamiltonian (\ref{eqn:Hamiltonian})), we globally address the ions using two off-resonant $\lambda=355$ nm laser beams (which we call $R1$ and $R2$) to drive stimulated Raman transitions \cite{Hayes2010,Campbell2010}. At the ion chain, the beam $R1$ with frequency $\omega_L$ perpendicularly intersects a multi-colored beam $R2$ with frequencies $\omega_L+\omega_S\pm\mu$.   Their wavevector difference $\Delta\vec{k}$ points along the $x$-direction of transverse ion motion and their frequency differences couple near the upper and lower $x-$motional sidebands. This configuration generates a spin-dependent force at frequency $\mu$ \cite{MolmerSorensen} and gives the Ising couplings \cite{QSIMPRL2009}
\begin{equation}
\label{eqn:Jij}
2\pi J_{i,j}=\Omega_i\Omega_j\frac{\hbar(\Delta\vec{k})^2}{2M}\sum_m\frac{b_{i,m}b_{j,m}}{\mu^2-\omega_m^2}
\end{equation}
in the Lamb-Dicke limit when the frequency $\mu$ is sufficiently far from the normal mode frequencies $\omega_m$. In Eqn. \ref{eqn:Jij}, $\Omega_i$ is the Rabi frequency at the $i^{\text{th}}$ ion, $M$ is the single-ion mass, and $b_{i,m}$ is the normal mode transformation matrix element for the $i^{\text{th}}$ ion in the $m^{\text{th}}$ mode. We set $\mu$ so that all $J_{i,j}>0$, resulting in long-range AFM Ising interactions that fall off with distance as $\sim 1/|i-j|^\alpha$, where $\alpha\approx 0.9-1.3$.

To apply the transverse field part of the Hamiltonian (second term in Eqn. \ref{eqn:Hamiltonian}), we add an additional component at frequency $\omega_L+\omega_S$ to the multi-color laser beam $R2$. The beatnote difference between $R1$ and this component of $R2$ drives carrier Rabi oscillations between the spin states $\ket{0}_z$ and $\ket{1}_z$, generating an effective magnetic field. We orient the field transversely to the spin-spin couplings by setting the phase of the component at $\omega_L+\omega_S$ equal to the average phase of the two components at $\omega_L+\omega_S\pm\mu$.

The amplitudes, frequencies, and phases needed to apply the Ising Hamiltonian are imprinted on the $\lambda=355$ nm laser beams using acousto-optic modulators (AOMs) driven by an arbitrary waveform generator (AWG). The AWG (Agilent M8190A) is programmed to output a voltage of form
\begin{eqnarray}
\label{eqn:AWG}
V(t)=&V_1&\sin[(\omega_{A}-\mu)t]+V_2\sin[(\omega_{A}+\mu)t+\varphi]\\
&+&V_3(t)\sin[\omega_{A} t+\varphi/2] \nonumber
\end{eqnarray}
where $V_1$ and $V_2$ are the amplitudes of the components that generate the $J_{i,j}$ couplings, $\omega_{A}$ shifts the frequency difference between $R1$ and $R2$ into resonance with $\omega_S$, and by our convention $\varphi$ is set to $\pi$ to define a spin-spin interaction $\sigma_x\sigma_x$. 
The time-dependent amplitude $V_3(t)$ determines the transverse field $B(t)$ and is made to decrease with a linear, exponential, or local adiabatic profile for these experiments. Because the phase of the carrier component $V_3(t)$ is the same as the mean phase of the two sideband components $V_1$ and $V_2$, the interaction is shifted by $\pi/2$ to give an effective magnetic field coupled to $\sigma_y$, after accounting for the inherent $\pi/2$ phase lag between carrier and the sideband transitions. The rf AWG output signal (Eqn. \ref{eqn:AWG}) is amplified to deliver a peak power of \mbox{1.8 W} to a 50$\Omega$ AOM in the beam path of $R2$, generating frequency components relative to $R1$ at $\omega_S-\mu$, $\omega_S+\mu$, and $\omega_S$ with corresponding amplitudes set by $V_1$, $V_2$, and $V_3(t)$.

\FigureThree

\section{Preparing AFM Ground States}
\label{sec:preparing}

We now measure the ability for each of the ramp profiles in Sec. \ref{sec:AQS} to prepare our spin system into the ground state of Eqn. \ref{eqn:Hamiltonian} at $B=0$. For this measurement, we use $N=6$ ions and create AFM spin-spin interactions of the form $J_{i,j}\approx(0.77~\text{kHz})/|i-j|$. These long-range AFM interactions lead to a fully-connected, frustrated system as all couplings cannot be simultaneously satisfied. Nevertheless, the ground state of the system is easily calculable for 6 spins and is found to be a superposition of the two N\'eel-ordered AFM states, $(\ket{010101}+\ket{101010})/\sqrt{2}$.

Fig. \ref{fig:rampdata}(a) shows the probability of creating the AFM ground state when the transverse field $B(t)$ is ramped using linear, exponential, and local adiabatic profiles. The total ramp time $t_f$ is varied from 0 to 2.4 ms, with a new ramp profile calculated for each $t_f$. Each data point is the result of 4000 repetitions of the same experiment, with error bars that account for statistical uncertainty as well as estimated drifts in the Ising coupling strengths. In agreement with the predictions in Sec. \ref{sec:AQS}, the data show that local adiabatic ramps prepare the ground state with higher fidelity than exponential or linear ramps. 

The solid lines in Fig. \ref{fig:rampdata} plot the theoretical prediction of the ground state probability with no free parameters. In each case we begin by numerically integrating the Schr\"odinger equation using Hamiltonian (\ref{eqn:Hamiltonian}) with the desired $B(t)$ and the initial state $\ket{\psi(0)}=\ket{000\ldots}_y$. At the end of the ramp, we calculate the overlap between the final state $\ket{\psi(t_f)}$ and the AFM ground state $(\ket{010\ldots}+\ket{101\ldots})/\sqrt{2}$ to extract the probability of the ground state spin configuration. We account for decoherence-induced decay of the ground state probability by multiplying the calculated probability at time $t$ by $\exp[-t/t_d]$, where $t_d$ is the measured $1/e$ coherence time of our spin-spin interactions. 

The fact that local adiabatic ramps do not yield $100\%$ ground state probability at $t_f=2.4$ ms is not surprising, given that the adiabatic condition is $t_f \gg 3.6$ ms for our experimental parameters. For comparison, the $\sim 80\%$ ground state population found with a 2.4 ms local adiabatic ramp would take an exponential (linear) ramp \mbox{9.7 ms} (29 ms) to achieve -- a factor of 4 (12) longer. However, these significantly longer ramps do not yield high-fidelity ground state preparation in practice, since significant decoherence effects arise in our experiment after about \mbox{2.4 ms}. Local adiabatic ramps therefore offer the best way to prepare the ground state with high probability.
 
The data in Fig. \ref{fig:rampdata}(b) show how the ground state probability grows during a single 2.4 ms linear, exponential, or local adiabatic ramp. The ground state population grows quickly under local adiabatic evolution since the transverse field $B(t)$ is reduced quickly at first. In contrast, the linear ramp does not approach the paramagnetic to AFM phase transition until $\sim 2$ ms, and the AFM probability is suppressed until this time. Once again, local adiabatic ramps show the largest ground state probability at each time.

\section{Determining Ground States}
\label{sec:determining}
Finding the ground state at the end of an adiabatic quantum simulation presupposes that the transverse field $B(t)$ is ramped adiabatically \cite{FarhiAQC}. However, as demonstrated in Sec. \ref{sec:preparing}, it can be difficult in many instances to satisfy the adiabatic criterion while avoiding decoherence effects, particularly in frustrated, fully-connected systems. In this section, we show that the ground state spin ordering may be extracted even when the ramp is non-adiabatic.

\FigureFour

To accomplish this goal of ground state identification, we examine the probability distribution of all spin configurations and select the most prevalent state. Consider an experiment where the spins are initialized into $\ket{000\ldots}_y$ (as usual) and the transverse field $B(t)$ is instantly switched from $B=B_0$ to $B=0$. Measurement along the $x$-direction would yield an equal superposition of all spin states; in this instance, the ground state is just as probable as any other state. If the transverse field $B(t)$ is instead ramped at a fast but finite rate, the quantum simulation is slightly more adiabatic than the instantaneous case, and the ground state becomes slightly more prevalent than any other state. When $B(t)$ is ramped slowly enough, the ground state population is nearly $100\%$ and dominates over that of any other state.

Using the single-ion resolution of our intensified CCD camera, we can directly measure the probability of creating each of the $2^N$ possible spin configurations. \mbox{Fig. \ref{fig:MostProbableState}} shows the measured probability for all of the 64 spin states at each local adiabatic ramp data point in Fig. \ref{fig:rampdata}(a). When the total ramp time is $0.00$ ms (i.e. instantaneous), we measure a distribution with nearly equal probability in each of the possible states, as expected. As the total ramp time is made longer (up to 2.4 ms), the populations in the two degenerate AFM ground states emerge as the most probable compared to any other spin configuration.

A close analogy may be drawn with a Landau-Zener process \cite{Zener} in a two-level system comprised of the ground and first coupled excited states. In the Landau-Zener framework, a system that starts in the state $\ket{000\ldots}_y$, the ground state of the Hamiltonian (Eqn. \ref{eqn:Hamiltonian}) when $B/J \gg 1$, will be transformed into the new ground state $\ket{111\ldots}_y$ at $B/J \ll -1$ if $B(t)$ is ramped adiabatically. Likewise, an instantaneous switch from $B/J \gg 1$ to $B/J \ll -1$ will leave the system in an excited state with $100\%$ probability.

Our experiment most closely resembles half of a Landau-Zener process, in which $B(t)$ starts with $B \gg J$ and ends at $B=0$. One can write an analytic expression to calculate the transition probability for this half-Landau-Zener evolution \cite{HalfLZ}, which has a maximum value of 0.5 for an instantaneous ramp. Any fast but finite ramp will give a transition probability $< 0.5$, and the ground state will always be more prevalent than the excited state.

The technique of identifying the most prevalent state as the ground state is subject to some limitations. First, the initial state (before the ramp) should be a uniform superposition of all spin states in the measurement basis -- a condition satisfied by preparing the state $\ket{000\ldots}_y$ and measuring along $\hat{x}$. If some spin states are more prevalent than the ground state initially, then some non-zero ramp time will be necessary before the ground state probabilities ``catch up'' and surpass these initially prevalent states. Second, the ramp must not cross any first-order transitions between ordered phases, as non-adiabatic ramps may not allow sufficient evolution time towards the new ground state order.

In addition, a good determination of the ground state requires that the difference between the measured ground state probability $P_g$ and next excited state probability $P_e$ be large compared with the experimental uncertainty, which is fundamentally limited by quantum projection noise $\sim1/\sqrt{n}$ after $n$ repetitions of the experiment \cite{QuantumProjectionNoise}. This implies that the most prevalent ground state can be determined reliably after repeating the measurement $n > (P_g^2+P_e^2)/(P_g-P_e)^2$ times.  Assuming an exponential distribution of populated states during the ramp (as may be expected from Landau-Zener-like transitions), the number of required runs should then scale as $n \sim (\bar{E}/\Delta)^2$ in the limit $\bar{E}\gg\Delta$, where $\bar{E}$ is the mean energy imparted to the spins during the ramp, and $\Delta$ is the energy splitting between the ground and first coupled excited state. 

If the gap shrinks exponentially with the number of spins $N$ (i.e. $\Delta\sim e^{-N}$), ground state identification requires an exponential number of measurements $n$ in the simulation. However, in cases where the gap shrinks like a power law ($\Delta\sim N^{-\alpha}$), the most prevalent state can be ascertained in a time that scales polynomially with the number of spins.  Regardless of the scaling, techniques that improve the ground state probability (such as local adiabatic evolution) can greatly increase the contrast of the most prevalent state and reduce the number of necessary repetitions.

\section{Scaling To Larger \textit{N}}
\label{sec:scaling}
In Secs. \ref{sec:preparing} and \ref{sec:determining}, we showed that local adiabatic evolution could improve ground state preparation and identification in a system of $N=6$ ions. As the system size increases, creating the ground state with high probability becomes much more difficult. However, we demonstrate that identification of the ground state remains robust in a system of up to $N=14$ spins using the most-prevalent-state selection technique.

\FigureFive

As the system size $N$ grows larger, two effects contribute to a shrinking critical gap $\Delta_c$, further reducing the adiabaticity transverse-field ramps in our frustrated AFM system. The first is the well-known result for transverse-field Ising models that $\Delta_c\rightarrow 0$ as the system size approaches the thermodynamic limit $N\rightarrow\infty$ \cite{SachdevBook}. The second effect arises from increasingly longer-range interactions at larger $N$ (see Fig. \ref{fig:Niondata}(a)) that lead to more frustration and smaller energy gaps in the system \cite{QSIM2013Science}. The combined effect is shown in Fig. \ref{fig:Niondata}(b), where the resulting critical gap for our experimental parameters decreases by a factor of 6 when $N$ is increased from 3 to 10 ions. The grey dotted curves in Fig. \ref{fig:Niondata}(b) demonstrate that even if the interaction range $\alpha$ is held fixed, $\Delta_c$ decreases with $N$ on account of the first effect.

Fig. \ref{fig:Niondata}(c) shows the probability of preparing the ground state using linear, exponential, and local adiabatic ramps as $N$ is increased from 2 to 10. At $N=10$, the fidelity falls to only $21\%$ for local adiabatic ramps, which is small but markedly better than exponential ($9\%$) or linear ($3\%$) ramps. Stronger $J_{i,j}$ couplings (which scale quadratically with increased $\lambda=355$ nm laser power) and longer ramp times (which would require a smaller rate of decoherence) are likely needed for high-fidelity adiabatic ground state preparation at larger $N$.

To show the potential scaling power of local adiabatic evolution, we perform quantum simulations with 12 and 14 ions (Fig. \ref{fig:Niondata}(d)). In this regime, we are unable to directly calculate the local adiabatic ramp profile using a standard desktop computer due to the exponential growth of the computation time (just building a $2^{14}\times 2^{14}$ matrix of machine-sized numbers requires over 2 GB of RAM). Instead, we approximate the gap $\Delta(B)$ by the piecewise function
\begin{equation}
\label{eqn:DeltaPiecewise}
\Delta(B)=\left\{
\begin{array}{rl}
\Delta_c~~~~~~~ \text{ if } B \leq B_c \\
\Delta_c+4(B-B_c) \text{ if } B > B_c
\end{array} \right.
\end{equation}
with $B_c$ and $\Delta_c$ extrapolated from the 3-10 ion calculations. This $\Delta(B)$ is then used to solve the differential equation \ref{eqn:gamma}. For $N\leq10$ the approximate local adiabatic ramp performs as well as the exact ramp to within experimental error, while for $N>10$ it continues to outperform exponential and linear ramps.

\FigureSix

Although the ground state probability becomes small for increasingly large $N$, the ground state spin ordering remains distinctly the most prevalent spin configuration even for $N=14$. Following the technique outlined in \mbox{Sec. \ref{sec:determining}}, we experimentally measure the probability distribution of creating each of the $2^{14}=16384$ possible spin states at the end of our quantum simulation. The two most prevalent spin states, the camera images of which are shown in Fig. \ref{fig:ProbsFourteen}(a), are again revealed to be the N\'eel ordered AFM states.

Fig. \ref{fig:ProbsFourteen}(b) demonstrates the resiliency of most-prevalent state selection to ramps that are far from adiabatic. Identification of the ground state proceeds easily, even though the total ground state probability is only $\sim 3\%$. The requirement of satisfying the adiabatic criterion (Eqn. \ref{eqn:adiabatic}) is replaced only by the requirement that the most prevalent state probabilities are accurately resolvable compared with those of any other states. While the method should remain robust for even larger $N$, more adiabatic ramps (generated by longer ramp times or stronger spin-spin couplings) will decrease the number of experimental repetitions needed to clearly resolve the state probabilities.

\section{Conclusion}
\label{sec:conclusion}
In conclusion, we have used local adiabatic ramps to prepare ground states with high probability in a trapped-ion adiabatic quantum simulator, as well as identify ground states in a system of up to 14 fully-connected spins. Local adiabatic ramps are found to maximize the ground state population compared with other adiabatic methods and require knowledge of only the lowest $\sim N$ energy eigenvalues of the Hamiltonian under study. As $N$ grows large and even the lowest eigenvalues are difficult to calculate, we have demonstrated that a simple, approximated local adiabatic ramp can still be used to improve the ground state preparation. We have additionally described a technique to determine the ground state spin ordering even when ramps are severely non-adiabatic, and have experimentally found the correct ground state in an $N=14$ frustrated AFM spin system. The technique should scale in principle to $N=30$ spins and beyond, where finding the ground states of complicated many-body spin systems becomes classically intractable.

\bigskip

We thank Jim Freericks for helpful discussions. This work is supported by the U.S. Army Research Office (ARO) Award W911NF0710576 with funds from the DARPA Optical Lattice Emulator Program, ARO award W911NF0410234 with funds from the IARPA MQCO Program, and the NSF Physics Frontier Center at JQI.

\bibliographystyle{prsty}
\bibliography{qsimrefs}
\end{document}